# The Office of the Future: Virtual, Portable and Global

**Jens Grubert**
Coburg University of Applied Sciences and Arts
jg@jensgrubert.de

**Eyal Ofek**
Microsoft Research Redmond
eyalofek@microsoft.com

**Michel Pahud**
Microsoft Research Redmond
mpahud@microsoft.com

**Per Ola Kristensson**
University of Cambridge
kristensson@acm.org

**Editors**:

Frank Steinicke

frank.steinicke@

uni-hamburg.de

Christian Sandor

chris.sandor@gmail.com

Virtual Reality has the potential to change the way we work. We envision the future office worker to be able to work productively everywhere solely using portable standard input devices and immersive head-mounted displays. Virtual Reality has the potential to enable this, by allowing users to create working environments of their choice and by relieving them from physical world limitations such as constrained space or noisy environments. In this article, we investigate opportunities and challenges for realizing this vision and discuss implications from recent findings of text entry in virtual reality as a core office task.

Much of the hype around Virtual Reality (VR) has focused on immersive gaming and entertainment, and considerable progress has been made in those directions in recent years. The underpinning thesis in this article, however, is that recent VR research progress allows us to also reimagine the office work of the future[1]. Raskar et al. imagined novel use cases for office work based on projection-based Augmented Reality, allowing local office workers with remote groups. Immersive head-mounted displays build upon this idea without the need for instrumentation of the environment with projector-camera systems and, hence, enable novel office experiences on the go. VR office based on immersive head-mounted displays (HMDs) open up a novel design space with exciting new opportunities for immersive, flexible and fluid office work.

Despite the rapid rise of mobile devices such as smartphones and tablets, the traditional workstation and laptop setups still dominate today's office work. Users type text on full-sized physical QWERTY keyboards and use a mouse or trackpad to select and manipulate on-screen objects. Common activities such as typing, editing text, changing the input focus between text fields, switching between windows in an application, and switching between applications use well-established keyboard shortcuts and direct manipulation techniques. Also, in stationary work settings, workers often use multiple-screens to create a larger display area. Past research indicates that large monitors enable more efficient work[2].

Supporting the above and other typical office activities in a VR environment requires translating the processes of familiar everyday office work practices into efficient and comfortable interaction techniques that simultaneously maximize the advantages posed by VR and minimize its limitations. A further constraint is path dependency: the tendency of users to prefer well-established processes despite being suboptimal in order to minimize learning effort.

# A VISION OF VR OFFICE WORK

VR headsets can filter users from the physical world and provide full control of the inputs to their senses, such as visual, auditory and haptics. This provides several advantages:

## Control of the environment around users

Many times, the physical environments surrounding users are clearly suboptimal. The available physical, as well as display space, might be small, and illumination may be less than adequate, resulting in a slew of disturbances all around users. An extreme example might be a person trying to work while sitting in an economy seat on an airplane (Figure 1).

Using VR head-mounted displays (HMDs), users can work in ideal environments of their liking: wide, well illuminated, private, and with a wide display area without outside disturbances.

## Location-independent repeatability of user experiences

Users who travel frequently might like to keep their familiar work environment constant (for example, the number of monitors, their order and arrangement of the applications around them, the shape of the room, notes on a virtual whiteboard, etc.) even when they are in different places with different physical constraints. This reduces context switching overhead and enable the use of muscle memory by the user during travel: as long as there is an access to a table to place a keyboard and a mouse, laptop or slate, users can carry a large virtual office with them wherever they go.

Virtual displays can recreate a similar arrangement of resources around the user in any location. Even if the recreated VR arrangement may be limited by the physical environment, due to for instance the lack of reachability or real-world haptics, it is possible to identify a VR arrangement that approximates the original one and leverage users' familiarity.

A VR office allows everyday office interactions to transition from locations to temporal events. Interactions can be accessed by temporal events. Instead of a meeting being accessed by presence in a dedicated meeting room, the meeting can continue from a snapshot of the moment where the last meeting ended: writing still appearing on the whiteboard and all relevant documents being open.

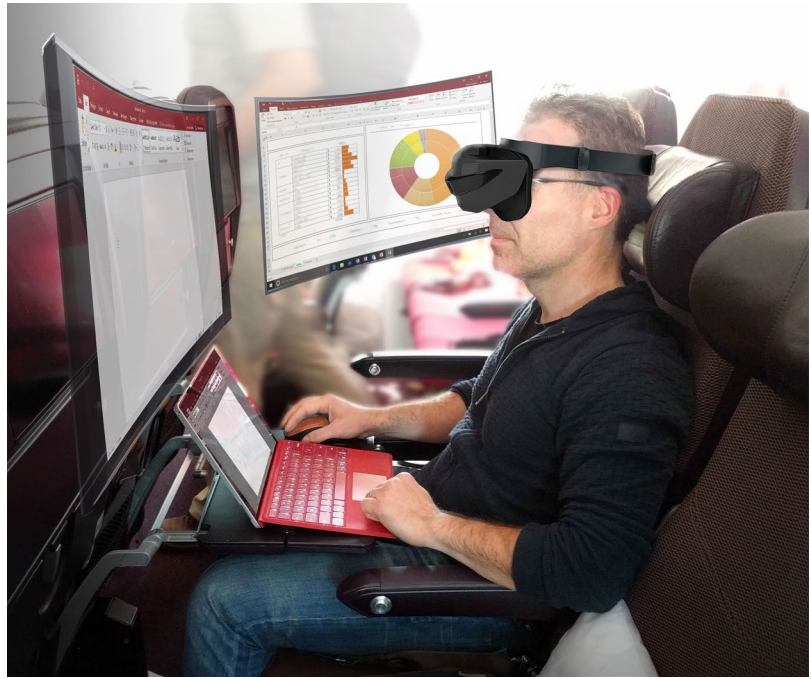

Figure 1: A virtual office environment in VR. A user may enjoy a large multi-display environment and background disturbance reduction, even in challenging environments.

## Privacy

Working in public environments exposes the contents of users' screens to unauthorized viewers in her vicinity. Directional visibility filters may lower visibility for people sitting next to the user but do not block all directions, such as people standing behind the user.

HMDs are personal and enable users to work without such privacy implication. Potential access to content can be controlled by the user.

However, privacy is still not fully guaranteed as onlookers, for instance could observe the user's typing. This opens up interesting research questions on mitigation strategies, such as introducing people around the VR user as avatars or mixed-reality blending of the surroundings with a virtual office.

## Relieve physical world limitations

The virtual world allows users to do things that are impossible in the physical world. They may move their hands and reach longer distances than their physical hand reach, change their appearance or draw on a whiteboard in front of them while their physical hands are resting on a table, reducing fatigue. Users may travel immediately to a meeting room somewhere else in the world. In the virtual world, there is a potential to equalize differences that may limit users from local resources, distances, or physical capabilities.

# CHALLENGES AND OPPORTUNITIES FOR OFFICE WORK IN VR

The above vision may hold multiple benefits, yet there are many challenges and technological improvements that need to be addressed to make VR-based office work practical for the public.

## Head-Mounted Display Quality

The field-of-view of current HMDs is substantially smaller than a human's field-of-view. The common horizontal field-of-view is around 100 degrees, which is about half of the natural field-of-view, and the vertical view angle is even smaller. Several upcoming HMDs offer 200 degrees horizontal field-of-view and in a few years, we may see HMD that will cover the full field-of-view of the user. Furthermore, the resolution of the HMD display is limited by the need to cover a very large view angle. Currently, this resolution is too low to make it effective for users to read small text, as is possible on a high-resolution monitor. Therefore, current VR applications use larger font sizes, which undermines the VR advantage of a large field-of-view. These limitations will probably be mitigated when new HMDs are introduced.

Another concern is that most of today's HMDs generate 3D images through stereoscopic image generation (i.e. generating separate 2D images for the left and right eye) resulting in vergence-accommodation conflicts that can have a negative impact on the user experience and performance in VR[3]. New technologies, such as lightfields or holographic displays may enable more natural views but have yet to reach consumer product levels.

Most current HMDs are tethered and use external sensors/beacons for tracking, limiting the user to a small volume of operation. This obviously results in a nonmobile VR setup. Although much of office work might be limited to around a desk area, there is an advantage in allowing users free movement without being restrained by wires, or coverage of room-based sensors. Again, there are already some commercial products that offer inside-out optical tracking, which is independent of environmentally-located sensors, as well as wireless transmission of VR content. Further challenges arise from using inertial-based tracking systems in mobile contexts such as cars[4].

Finally, any error in the tracking of the user's motion or latency in the reaction of the display content to the user motion may increase the risk for the generation of motion sickness[5]. The nauseating feeling raises from a disagreement between the user senses, mainly the visual one and the vestibular system that monitors our balance.

## Situational awareness and physical isolation

VR is at one extreme end of the reality-virtuality continuum. This can be beneficial as a user is potentially more immersed in the task at hand and it is plausible this could have positive ancillary effects, such as better concentration and less stress due to the removal of distractions in the environment. On the other hand, VR may also result in a loss of situational awareness and lead to unwanted physical isolation. The current popular applications are entertainment-oriented, and as such, they tend to use the immersive nature of the VR display to replace the user environment with a new one and give the impression of being in a different reality.

The use of VR in a work environment may be a mix use of both reality blocking (removing disturbing elements, having larger screens, etc.) as well as environment representation, enabling manipulation of physical objects, environmental awareness, and communication. Current approaches to move the operating point on the reality-virtual-ity continuum and use mixed reality to maintain a connection to the physical surroundings ranges from streaming stereo video of the environment to the display (video-based AR) to modeling the environment and representing it in the virtual world[6]. This opens up a rich design space. In this context, an open research question is whether there are any situational awareness or physical isolation issues in VR office work, and if so, how these effects could be quantified and understood in terms of contributing factors. Such investigations can help to identify design principles for future systems.

## Fluidity, flow and locus of control

Users' sense of agency and locus of control is an indicator of usability, as evidenced by its inclusion in user interface guidelines and research on agency. It is unclear how VR affects users' sense of control of their own actions. Related, flow can be important for office work. It is also unclear whether VR office work is likely to increase or decrease flow.

It is an open question whether effective mitigation strategies that minimize loss of positive VR office work benefits can be identified. It is possible to envision several strands of research, including investigating the relative effects of video-based mixed reality vs. optical see-through augmented reality[7]. It may also be interesting to explore minimal interventions in the form of some type of awareness-markers that relate to the physical surroundings which can be subtly introduced in the VR environment. The translation of such awareness-markers to VR need not be graphical, but could also use audio cues or haptic feedback.

## Communication between users

The need to wear an HMD blocks the view of the user's face from the environment, resulting in loss of an important communication channel between people. Although, recent research attempt to recover this channel by methods ranging from virtual avatars representing the users and their facial expressions, internal sensing within the headset, to using prior captured data to better synthesize the view of the user's face. Currently, this is an active field of research.

## Typing and control efficiency

A key challenge is to minimize the performance gap between ordinary office work, in particular, typing and editing, using a workstation or laptop setup vs. a VR setup. Typing is a learned motor skill and recent empirical research has discovered users can be clustered into a small set of different typing styles and type using their own full-sized keyboards at an average rate of 52 words-per-minute, where a word is defined as five consecutive characters including spaces[8].

In addition to typing text, users also spend considerable effort editing text. This requires interaction techniques that are both fast and precise, which can be challenging if the input is relying on noisy sensor data, such as depth sensing. In contrast, established mice and touchpads provide users with robust control, but at the expense of being 2D input devices that can be challenging to use for 3D interaction. Still, text editing on a standard PC is typically conducted using mouse and keyboard and we will indicate later in this article that text entry in VR can also benefit from standard keyboards.

Furthermore, virtual environments, unbounded by the limitation of a physical world, can introduce new interaction techniques that may prove to be even more efficient than current physical ones. For example, a physical keyboard is limited to lie on a supporting surface such as a table, far away from the display and the edited document. This distance results in large head movement for occasional glancing at the keyboard, slowing down the work and may generate back and neck pains. In contrast, the virtual keyboard and the user's hands can be remapped from their physical locations to positions closer to the edited documents. Another example involves changing the look and transparency of the user's hands to enable better visibility of the keyboard and the edited document during manipulation (for example, see Figure 3), bottom-left. The design space of such possible alterations of reality is vast.

## TEXT ENTRY IN VR

The main focus of our research so far has been on text entry, as it is fundamental to many tasks ranging from document editing to internet browsing, and a task that has a considerable learning curve (most users are not fluent in touch-typing, and still use a various hunt-and-peck and other improvised strategies[7]). In fact, the cost of learning this task has prevented much progress of keyboard technology since the introduction of mechanical typewriters. Among most users, a combination of a traditional keyboard and a large, high-resolution monitor is still the preferred input method for editing longer text documents, working on spreadsheets or form-filling activities. Given the above observation, we set out to leverage user familiarity with traditional keyboards, and the widespread of such off-the-shelf devices, for work in VR, while using VR freedom of the physical world to improve the user experience.

Initially, it is not obvious that existing physical keyboards or nowadays common touchscreen keyboards are suitable for typing in virtual reality. The wearable displays block users' view of the real world, including their physical hands and the keyboards, either physical or touchscreen-based, and create a challenge to appropriately represent them in the virtual world. Despite today's VR HMD-limitations, we believe that the ability to control the user's environment, generating virtual displays that are as large as needed, both flat and three dimensional, the flexible mapping of the user's interaction space to the virtual space, and the advantage of privacy, may eventually make VR HMDs suitable environments for text entry and document editing. To investigate the potential of today's available hardware, we have studied text entry using standard keyboards (using the QWERTY layout), as described next.

Our user study of the performance of typing on physical and touchscreen keyboards[9] revealed that, while a user's typing speed in a baseline virtual environment is markedly slower than typing in the physical environment, users typed at an average of 60% of their usual typing rate when working in VR. We attribute this loss of speed to two factors: 1) the novelty of the setup and user's lack-of-experience with VR; and 2) the limitations of today's VR HMDs (specifically, lower resolution and latency). A key finding, however, is that typing skills transfer seamlessly from the real world to the virtual world.

VR allows the system to situate the keyboard wherever and whenever needed based on context; for example, placing it closer to the document or object of interest, and displaying a graphic representation of the user's hands in relation to the keyboard (in our experiments, we used circles representing the fingertips), see Figure 2. While this eliminates the need to constantly shift attention between the keyboard and document, it may also require the users to reposition their hands while typing. While such repositioning of the keyboards and hands proved to have little impact on typing efficiency with a physical keyboard, it resulted in some degradation of performance on touchscreen keyboards (perhaps due to the change of the direction of the finger motion as they disconnect from the touch surface).

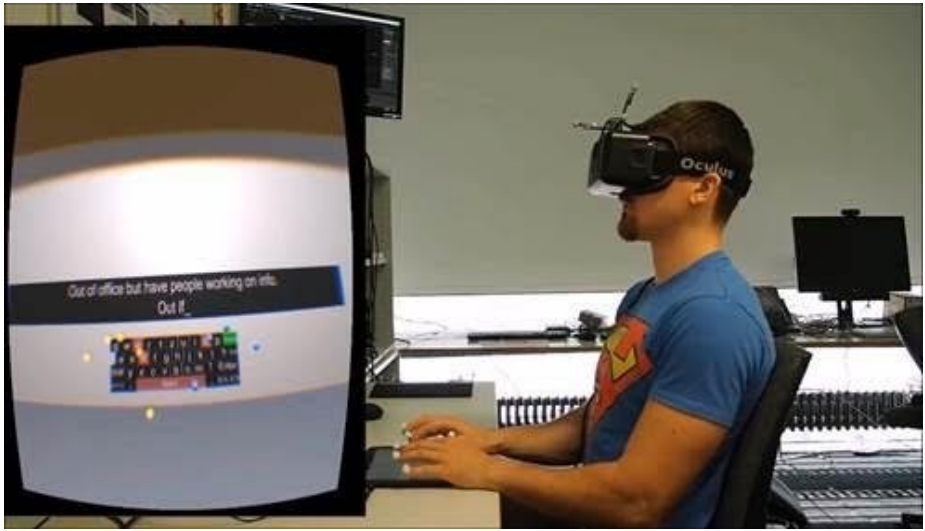

Figure 2: Displaying the user's hands in the view direction, rather than at the natural position has the potential to help the user remain focused on the document. It also has little to no impact on typing performance when using a traditional keyboard.

Another freedom VR provides is changing the representation and display of the user's hands in the virtual environment[10]. For example, the user's hands can become translucent in the virtual environment, which might provide an unobstructed view of the keyboard.

We presented users with four different hand representations as they typed in a VR scene (see Figure 3). The first two methods were analogous to traditional input methods; the third and fourth methods used manipulations only possible in VR:

1. A video of the user's hands, which is closest to the natural situation of typing without VR. However, the quality of such video is depended on the conditions of the physical environment, and it may limit the manipulations that can be generated in the virtual world, such as movement of the hands in space.

2. A full 3D model of the users' hands animated according to the tracking of the user's real hands.

3. A minimalistic 3D model in which most of the users' palms were transparent, and only the users' fingertips were displayed, to maximize the visibility of the keyboard.

4. Only showing the keys being pressed on the keyboard; that is, with hands that are completely transparent.

Surprisingly, the minimalistic model of the transparent hand with only fingertips visible was as easy to use and as efficient as blending a video of the users' hands. Such a model is easy to animate (it only requires sensing of the user's fingertips), and as a 3D model, it supports a large variety of manipulations in the virtual space. In contrast, the full 3D model of the hand was not as useful; subtle differences in the model's motions, as well as differences between the look of the model and the actual look of the user's hand, may have generated a dissonance between the user and the model and thereby reduced typing speed and accuracy. In fact, the results of the full 3D model were as poor as not revealing the hands at all to the user.

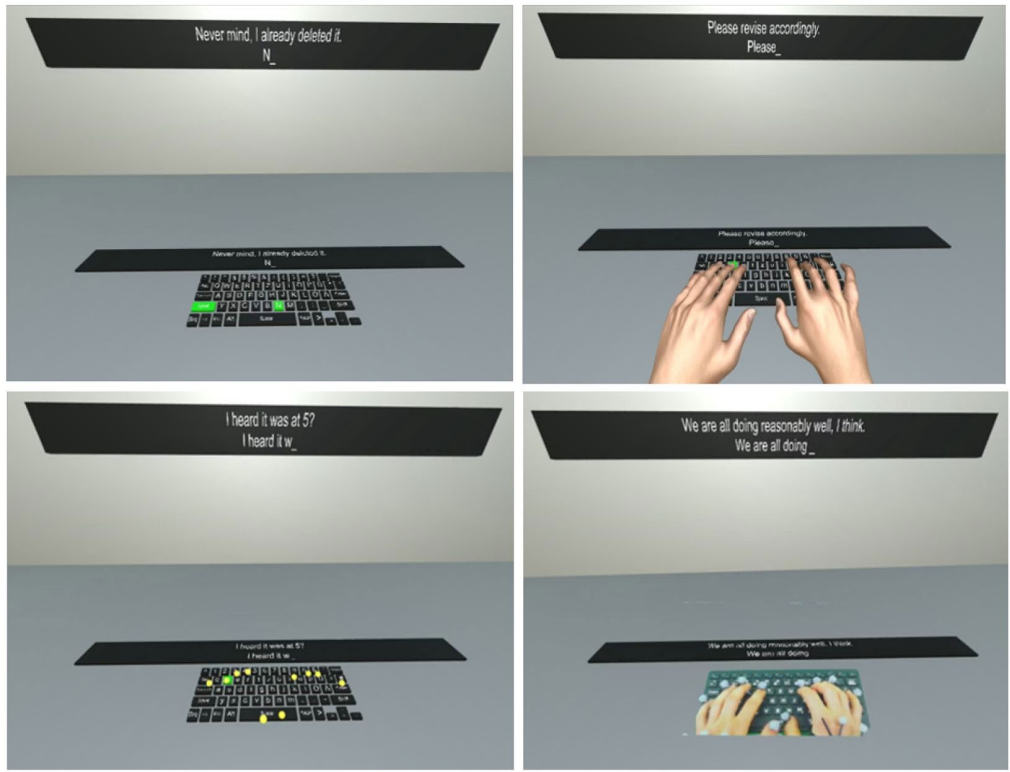

Figure 3: Clockwise from top-left: no hands, an inverse-kinematic hand model, a video blending of the user's hands, and fingertips as spheres.

## BEYOND CURRENT OFFICE TASKS

Text entry and document editing is an important and common task of today's office work, yet many other tasks could potentially benefit from the VR medium. For example, meetings can be independent of distances, travel time, and availability of meeting rooms and their instrumentation. Conversations, recorded by wearable microphones are easier to transcribe and translate, people, objects and social happenings in virtual spaces can be easier to analyze and describe to people that cannot visually observe the meeting room. Conversations may be mediated to include relevant information or help people challenged in social situations by using the private display of each participant[11] and more, see Figure 4.

Even more exciting might be the opening up of new opportunities that are impossible today, or are limited in their reach. In a VR office, there is practically no importance for the physical location of the users. It may open up jobs for remote people or people with disabilities that were prevented to join the workforce as equals. VR can enable people literally to see the work from other people's point of view, which may help communication and improve empathy, remote help, education, and reduce misunderstandings and disputes. VR has the potential to better use users' limited attention and mental resources, by minimizing travel and smoothing out transitions between tasks to minimize ramp-up costs, and control external disturbances based on the user's activities and estimated concentration. These are just a few possible future applications and potential benefits. We believe the freedom of the VR world along with very accurate sensing of users' movement, their attention, and behaviors will prove to be a fertile ground for more such transformative applications that eventually will reimagine the office work as we know it.

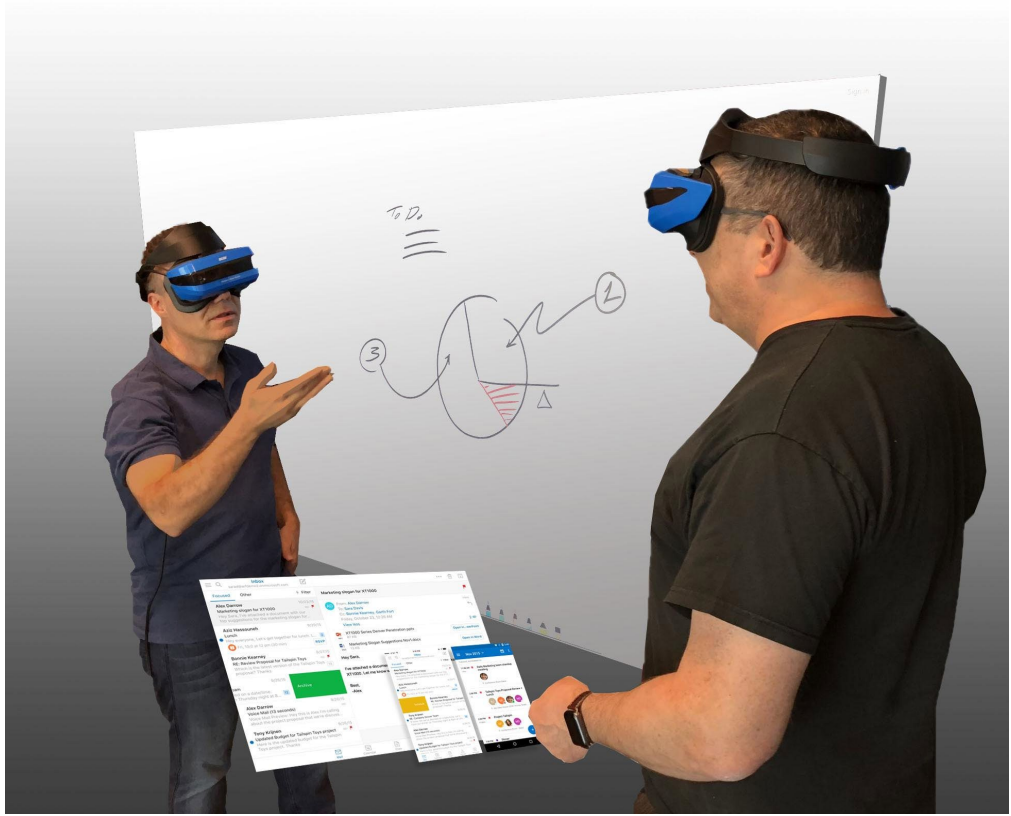

Figure 4: Private displayed content could support conversations, in particular for people challenged with social interaction.

AUTHOR VERSION

## ABOUT THE AUTHORS

**Jens Grubert** is Associate Professor for Human-Computer Interaction in the Internet of Things and lab director of the mixedrealitylab, a laboratory for Augmented and Virtual Reality, at Coburg University, Germany. He received his Dr. techn (2015) with highest distinction at Graz University of Technology, his Dipl.-Ing. (2009) with highest distinction at Otto-von-Guericke University Magdeburg, Germany. He is author of more than 50 peer reviewed publications and patents and published a book about Augmented Reality development for Android. His current research interests include interaction with multimodal augmented and virtual reality, body proximate display ecologies, around-device interaction, multi-display environments and cross-media interaction. Contact him at jg@jensgrubert.de.

**Eyal Ofek** is a senior researcher in the Redmond lab of Microsoft Research. His research interests include Augmented Reality, Virtual Reality, HCI and Haptics. He has a PhD in computer vision from the Hebrew University of Jerusalem. Contact him at eyalofek@microsoft.com.

**Michel Pahud** has a Ph.D. in parallel computing from the Swiss Federal Institute of Technology, Laussane. He won several prestigious awards including the Logitech prize for an innovative industrially-oriented multiprocessors hardware/software project. He joined Microsoft in 2000 to work on many different projects including videoconferencing/networking technologies and research in education. More recently, he has been focusing on human-computer interaction at Microsoft Research. His research includes bimanual interaction, novel form-factors, context sensing, haptics, augmented reality and virtual reality. Contact him at mpahud@microsoft.com.

**Per Ola Kristensson** is a University Reader in Interactive Systems Engineering in the Department of Engineering at the University of Cambridge and a Fellow of Trinity College, Cambridge. He is interested in designing intelligent interactive systems that enable people to be more creative, expressive and satisfied in their daily lives. He is an Associate Editor of ACM Transactions on Computer-Human Interaction and ACM Transactions on Intelligent Interactive Systems. Contact him at kristensson@acm.org.